# Drastic softening of Pd nanoparticles induced by hydrogen cycling

**Authors:** Jonathan Zimmerman[1], Maria Vrellou[2], Stefan Wagner[2], Astrid Pundt[2], Christoph Kirchlechner[2], and Eugen Rabkin[1]*

[1] Department of Materials Science and Engineering, Technion – Israel Institute of Technology, 3200003 Haifa, Israel

[2] Institute for Applied Materials (IAM-MMI and WK), Karlsruhe Institute of Technology, 76131 Karlsruhe, Germany

*Corresponding author. Email: erabkin@technion.ac.il

# Drastic softening of Pd nanoparticles induced by hydrogen cycling

**Abstract:**

Single crystalline faceted Pd nanoparticles attached to a sapphire substrate were fabricated employing the solid state dewetting method. The as-dewetted nanoparticles tested in compression exhibited all features of dislocation nucleation-controlled plasticity, including the size effect on strength and ultrahigh compressive strength reaching up to 11 GPa. Hydrogen cycling of as-dewetted Pd nanoparticles resulted in their drastic softening and in change of the deformation mode. This softening effect was correlated with the high density of glissile dislocations observed in the cycled particles. This work demonstrates that the nanomechanical behavior of hydride-forming metals can be manipulated by hydrogen cycling.

The Pd-H system is the longest- and best-studied metal-hydrogen couple forming the interstitial metal hydride [1]. In his pioneering work Graham has observed that a Pd foil that underwent repeated hydrogen cycling was "much crumpled and rather friable" [2]. Most likely, this was the first observation of strain hardening of a metal induced by phase transformation. Later it was confirmed that hydrogen cycling of bulk coarse-grain Pd results in an increase of dislocation density by up to three orders of magnitude and concomitant two-fold increase of yield stress from 100 to 200 MPa [3]. Also, the nanohardness of a coarse-grained Pd foil subjected to a full metal-hydride-metal cycle has increased by about 50% [4]. The increase in dislocation density is related to internal stresses in the metal matrix generated by the nucleation and growth of the hydride phase. Indeed, formation of the PdH hydride is associated with the linear lattice expansion in the range from $\varepsilon=2.8\%$ to 3.7% [1,5,6] which causes high stresses in the neighboring metallic matrix by far exceeding the elastic limit of bulk Pd [7]. The situation may be different in Pd samples of sub-micrometer size in at least one dimension because of the increase in strength and the change of plasticity mechanisms with decreasing sample size and constraint conditions [6,8,9]. Indeed, the maximum stress $\sigma$ hypothetically induced by Pd hydride formation can be roughly estimated as $\sigma \approx E\varepsilon \approx$ 4.5 GPa, where $E \approx$ 121 GPa is the Young modulus of polycrystalline Pd [10]. More elaborate finite element method calculations for Pd single crystal result in the values in the range of 4.5-5.5 GPa depending on the geometry of hydride precipitate and crystallographic direction [11]. The yield stress of comparable amplitude has been reported for Pd nanowhiskers grown by molecular beam epitaxy method [12], and for nanoparticles of several metals fabricated by solid state dewetting [13–15], so that it is well possible that small and defect free Pd samples can be fully transformed into the hydride phase without any plastic deformation. The question of whether the dislocation-mediated plasticity is activated during hydrogenation of small Pd samples has important thermodynamic implications: Fully elastic transformation with coherent metal/hydride interface results in increased pressure hysteresis and precludes the two-phase equilibrium during hydrogenation [16–18].

Several experimental studies indicated the possibility of fully coherent metal-hydride transformation in Pd nanoparticles and thin films in which the transformation strain is accommodated exclusively by elastic deformation, and the nucleating hydride phase swiftly sweeps through the sample precluding any two-phase equilibrium. Ulvestad *et al.* have reported in their *in-situ* Bragg coherent diffraction imaging (BCDI) study that dislocations are generated in the metal phase of partially hydrogenated Pd nanoparticles only in the particles larger than 300 nm in size, while the fully coherent transformation mechanism is likely in their smaller counterparts [19]. On the contrary, the critical size for fully coherent transformation of only 30 nm was reported in the study of Syrenova *et al.* performed with the aid of single-particle plasmonic nanospectroscopy [20]. This critical size is close to the thermodynamic estimates presented by Griessen *et al.* [21]. Syrenova *et al.* have also presented indirect thermodynamic evidence of incoherent transformation during reverse hydride-metal transformation in all studied particles [20]. A fully coherent metal-hydride interface was observed in the *in-situ* scanning tunneling microscopy and X-ray diffraction study of hydrogenation of Pd films less than 22 nm in thickness performed by Wagner *et al.* [22]. This critical thickness is suggested to be lowered compared to free samples by the adhesion to the rigid sapphire substrate, which results in stronger vertical expansion upon hydride formation and large in-plane stresses [23].

The short overview presented above indicates that there is still large uncertainty concerning the critical particle size for fully coherent metal-hydride transformation in Pd nanoparticles. Also, even in the case of fully coherent metal-hydride transformation it is unclear whether some dislocations are generated during reverse hydride-metal back-transformation upon hydrogen desorption.

The aim of the present work was to determine whether the hydrogen cycling of pristine, defect-free Pd nanoparticles of different sizes and shapes results in the change of their internal microstructure and formation of dislocations. To this end, we used the uniaxial compressive strength of Pd nanoparticles as an indicator of the presence of internal defects, primarily dislocations. Indeed, several studies indicated that ultra-strong metal nanostructures are very sensitive to ion irradiation and plastic pre-straining which result in their drastic softening [24,25].

The arrays of Pd nanoparticles attached to c-plane oriented polished sapphire substrate (University Wafers Inc.) were fabricated by the solid state dewetting method, as described in the previous works [13–15]. In short, the 10 nm-thick Pd film was deposited on thoroughly cleaned sapphire wafer employing the electron beam deposition method. The samples were annealed at a temperature of 1000 °C for 24 h in the flow of Ar-10 vol.% $H_2$ gas under a pressure slightly exceeding 1 bar and, thus, a hydrogen partial pressure of about 100 mbar. To avoid the formation of a hydride during dewetting heat treatment, the gas flow was switched to pure Ar at temperatures below 120 °C during heating and cooling of the sample. This temperature was selected based on the equilibrium bulk Pd-H phase diagram [1]: indeed, at the temperature of 120 °C the equilibrium plateau pressure of hydrogen corresponding to two-phase metal-hydride equilibrium is 320 mbar, significantly higher than the partial hydrogen pressure in the annealing gas of 100 mbar. This difference in pressures makes the formation of hydride phase during dewetting heat treatment highly unlikely. The samples were characterized using the following microscopes: For scanning electron microscopy (SEM), and electron backscattering diffraction (EBSD) - Ultra-Plus (Zeiss) field emission gun SEM fitted with an eFlash HD system (Bruker) EBSD detector, and Zeiss Merlin Gemini II field emission gun SEM. For transmission electron microscopy (TEM) - FEI / Thermo Fisher Titan Themis G2 60-300 transmission electron microscope (TEM) using an accelerating voltage of 300 kV in scanning transmission electron microscope mode (STEM). One of the samples has undergone four room temperature (21 °C) hydrogen absorption/desorption cycles of 1 h each under the hydrogen pressure of 1020 mbar during absorption and a background pressure of $3\times10^{-8}$ mbar in the UHV chamber. Even though the plateau pressure of hydride formation in adhered Pd nanosystems can be enhanced by up to two orders of magnitude [26] at room temperature, it is likely that the hydride formation pressure is exceeded and Pd nanoparticles formed a hydride phase under the applied hydrogen loading conditions. We note that for thin Pd films on sapphire substrates the hydride transformation was detected down to 10 nm film thickness at ambient conditions [6]. Also, the intrinsic inhomogeneity of hydrogen distribution in Pd nanoparticles can in fact lower the transformation pressure of the particles, as observed by Baldi *et al.* [27]. In the context of the present work this ensures full transformation under selected hydrogen cycling conditions. In the following this dewetted and hydrogen treated sample is denoted as "cycled".

The microcompression tests were performed with two nanomechanical testing instruments: Bruker PI85 Picoindenter installed at the Technion in Haifa and equipped with a 1 μm diamond punch, and Bruker PI89 Picoindenter installed at the Karlsruhe Institute of Technology (KIT) and equipped with a 2 μm diamond punch. All tests were performed in displacement-controlled mode, with a displacement rate of 1 nm/s. A strong consistency between the results obtained by using two different testing systems, testing at different locations and with three months in between the experiments demonstrates a high degree of data reproducibility [28,29].

The SEM image of the as-dewetted Pd sample is shown in Fig. 1a. The lateral particle size (defined as a square root of projected area of the particle) varies between 100 and 500 nm; a number of particles exhibit a top facet of two- or four-fold symmetry, while the top facet of some other particles represents a distorted hexagon. The EBSD measurements identified the former and the latter types of the particles with the (100) and (111) out-of-plane orientations, respectively (see Fig. S1 of Supplementary Material). The presence of two particle orientations is most likely related to abnormal growth of (100)-oriented grains in otherwise [111]-fiber textured film at the onset of dewetting [30]. A set-up of the microcompression test of a typical (100)-oriented particle and a side view of the particle before and after the test with the PI85 Picoindenter are shown in Fig. 1b. The vertical height of most as-dewetted particles is lower than their width, giving them a flattened appearance. It is interesting that the SEM images of the as-dewetted and cycled samples are nearly identical and do not show any change of the particle morphology (see Fig. S2 of Supplementary Material).

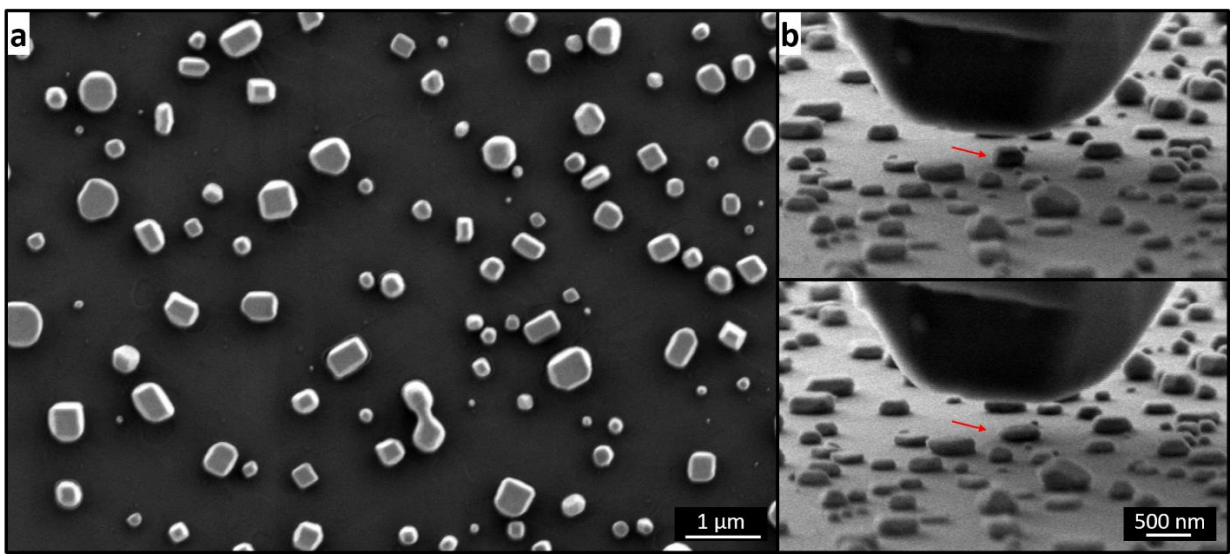

Fig. 1. The top-view SEM image of the as-dewetted sample with Pd nanoparticles (a). The side-view SEM image of the microcompression set-up and a typical tested (100)-oriented particle before (top) and after (bottom) compression (b).

Load-displacement curves of several typical (100)-oriented particles in the as-dewetted and in the cycled samples acquired with the PI85 instrument are shown in Fig. 2a. All tested particles in the as-dewetted sample exhibit a type of behavior typical of nucleation-controlled plasticity: a reversible elastic deformation followed by an abrupt and large displacement burst. On the contrary, the Pd particles in the cycled sample exhibit jerky load-displacement curves with small load drops sometimes preceded by minor displacement bursts. This is a typical behavior for the samples with significant initial dislocation population tested in displacement-controlled mode [24,25,28,31,32]. Thus, the load-displacement curves suggest the presence of initial dislocations in the cycled sample. The load-displacement curves obtained with the aid of PI89 instrument are in quantitative agreement with the results presented in Fig. 2a (see Fig. S3 of the Supplementary Material).

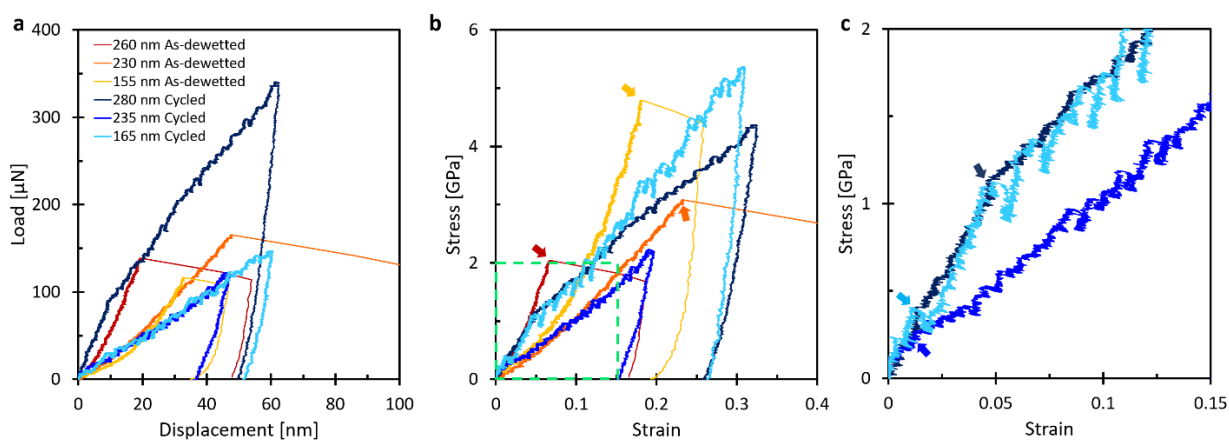

Fig. 2. The load-displacement (a) and engineering stress-strain (b) curves of several selected (100)-oriented Pd particles in the as-dewetted and cycled samples. (c) A zoomed-in area of low loads and strains of the cycled particles, see green dashed square in (b). The first plasticity-related events in the as-dewetted and cycled particles are marked with arrows in (b) and (c), respectively.

The engineering stress-strain curves obtained from the raw microcompression data by normalizing the load and displacement by the top facet area and initial particle height, respectively, are shown in Fig. 2b. Figure 2c shows a zoomed-in area of low stresses and strains showing how the strength of the cycled particles was determined: it was identified either with a first significant load drop, or with a marked change of slope [31]. One can clearly see that the first plasticity-related event in the cycled sample occurs at much lower values of stress as compared with the critical stress of the onset of plasticity in the as-dewetted sample.

The dependence of (100)-oriented particles strength on the diameter of their top facet is shown in Fig. 3. The strength of as-dewetted particles is defined as a critical stress at the onset of displacement burst exemplifying mass nucleation of new dislocations, whereas the strength of their cycled counterparts corresponds to the first plasticity-related event (see arrows in Fig. 2c). The strength of as-dewetted Pd particles exhibits a significant size effect, with the absolute values of strength being comparable to those of pristine particles of other face centered cubic metals and reaching a remarkable value of 11 GPa for the smallest particle [13–15]. The Pd particles in the hydrogen cycled sample exhibit a significantly lower yield strength than their as-dewetted counterparts, as shown in Fig. 3. The strength-diameter plot combining the data for (100)- and (111)-oriented particles exhibits a similar trend (see Fig. S4 of Supplementary Material). Performing a linear regression fitting of the data for the as-dewetted and cycled Pd particles yields the following relationship for a softening effect, $S$, defined as a ratio of yield strengths of as-dewetted and cycled Pd nanoparticles with the same top facet diameter, $D$:

$S = 98 \times D^{-0.588}$ (1)

where $D$ is measured in nm. Thus, for the particles studied in the present work the softening effect changes from 6.5 to 2.9 for the particles with the top facet diameter of 100 and 400 nm, respectively.

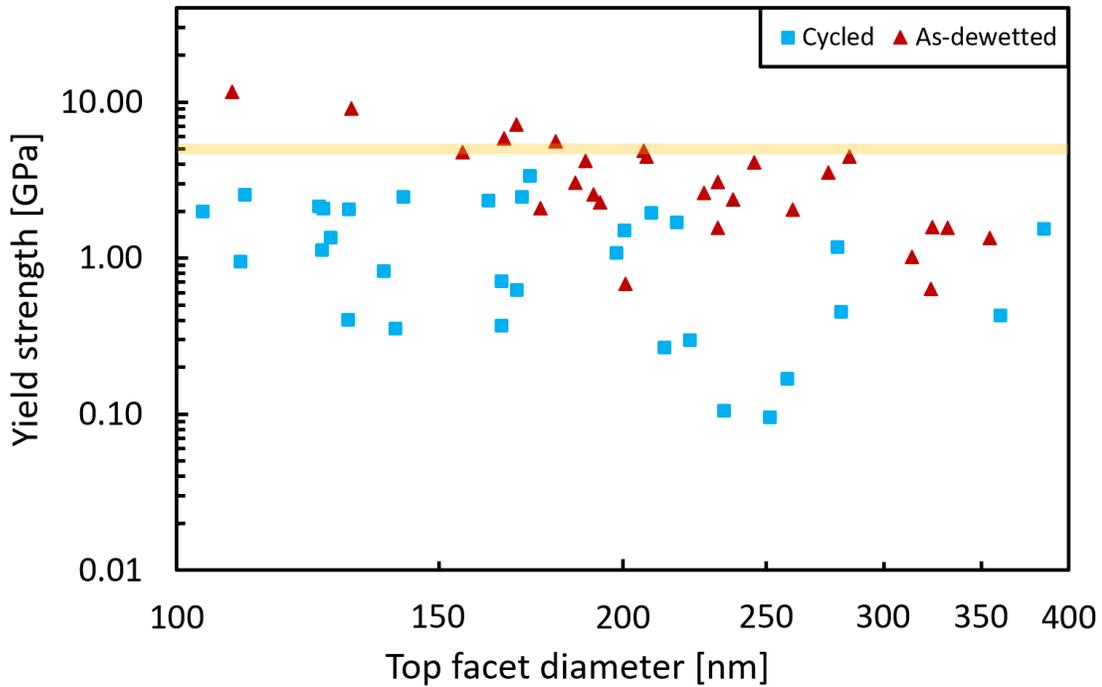

Fig. 3. The dependence of strength of the as-dewetted (triangles) and cycled (squares) (100)-oriented Pd particles on the size of their top facet, presented as a log–log plot. The strength of the particles is significantly lowered by hydrogen cycling. Further, the strength of the as-dewetted Pd particles exhibits a significant size effect. The horizontal line gives the range of stresses due to hydride transformation, showing that small particles yield above the transformation stress. These particles can be expected to transform coherently upon hydrogenation.

Cross-sectional TEM micrographs of representative Pd particles in as-dewetted and cycled samples prior to the microcompression test are presented in Fig. 4. The top facets of both particles are parallel to the substrate (Fig. 4a). The bright field image of the top facet acquired with higher magnification captures negligible surface roughness of the top facet in the as-dewetted particle, and a wavy character of the top facet with the perturbation amplitude of 1-2 nm in the cycled particle. This is an additional hint on dislocation activity during hydrogen cycling [33,34]. The STEM images of the near-interface and near-triple line regions reveal in the as-dewetted particle only small dislocation loops as they are often found in FIB-prepared cross-sectional samples [35]. The as-dewetted particle can be considered dislocation-free. Long dislocation lines are detected in the particle from the cycled sample (Fig. 4c). This demonstrates the presence of dislocations remaining in the cycled particles after hydrogen cycling. The dislocation density ρ in the cycled Pd nanoparticle determined using the method of Bailey and Hirsch [36] is $5.7\times10^{14}$ $1/m^2$. This value falls between ρ = $10^{13}$ $1/m^2$ in Au nanopillars after 35% of pre-straining [37], and ρ = $2\times10^{15}$ $1/m^2$ for the density of geometrically necessary dislocations in local pockets with high dislocation density in Cu micropillars after straining to 10% [38].

The results of our microcompression tests (Fig. 2) and microstructural data (Fig. 4) clearly indicate that as-dewetted Pd particles (even the larger ones as shown in Fig. 3) are free from line (dislocations) and planar (interfaces) defects, and their compressive strength is comparable to the theoretical strength of Pd [12]. Assuming a hydride transformation strain causing a maximum stress of 4.5–5.5 GPa, given by the horizontal yellow line in Fig. 3, small particles should exhibit coherent transformation into Pd hydride without any dislocation plasticity [9,16–18]. The data of Ulvestad *et al.* [19] and Baldi *et al.* [27] are consistent with this conjecture, especially taking into account the size effect on strength (Fig. 3). Thus, our data provide a quantitative support for the possibility of a coherent transformation mechanism in small Pd nanoparticles, since they clearly show that the strength of smaller particles is higher than the maximum stress associated with the metal-hydride transformation.

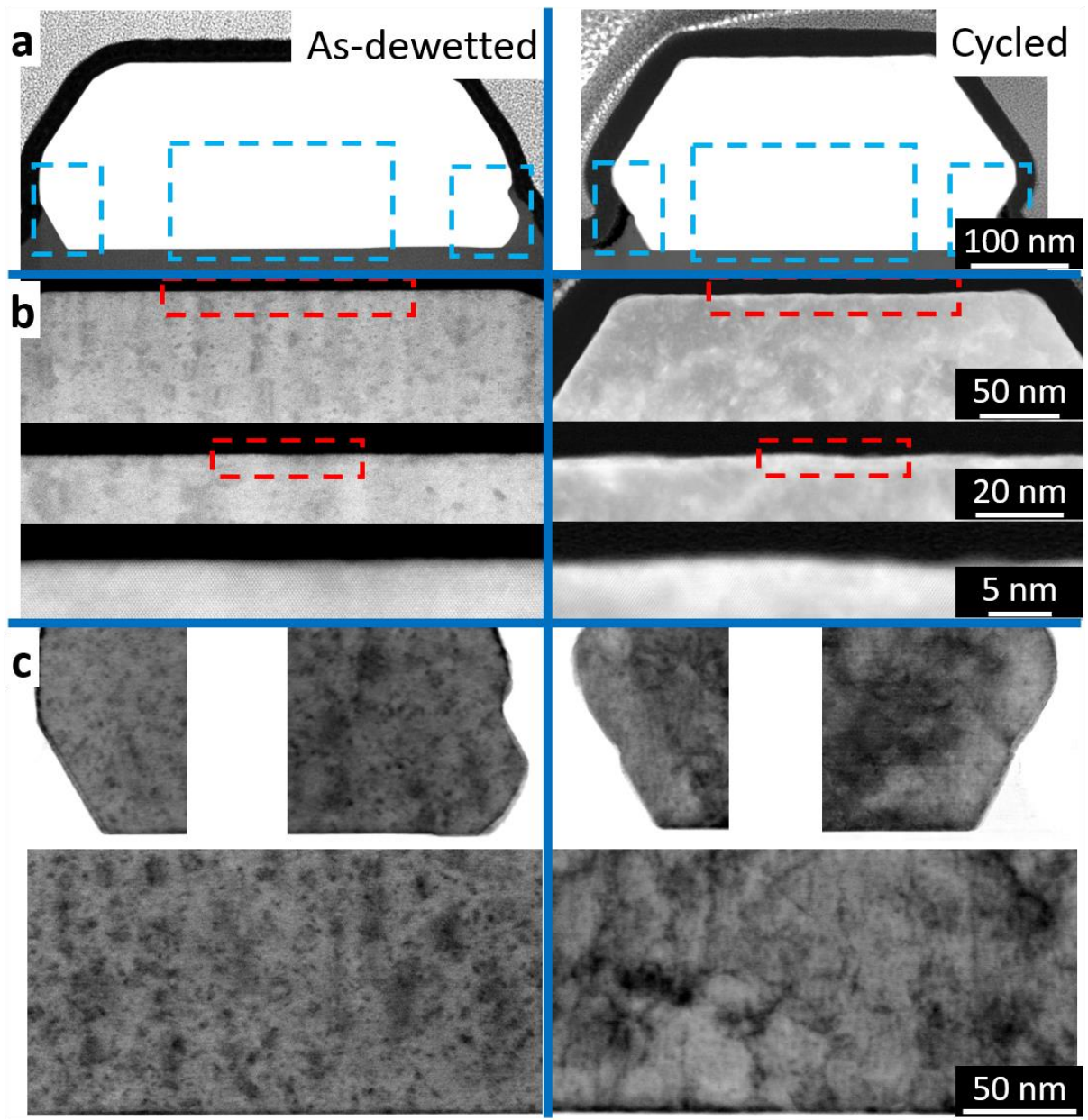

Fig. 4. Cross-sectional TEM micrographs of representative larger Pd particles in the as-dewetted and cycled samples, before the compression test. (a) general view presenting the particle morphology; (b) near-surface region demonstrating that surface roughness of the cycled particle is higher than that of its pristine counterpart; (c) STEM images of three near-interface regions demonstrating only small dislocation loops in as-dewetted particle originating from the FIB milling, but long dislocation lines in its cycled counterpart.

Our data on the cycled particles of all sizes unequivocally demonstrate the presence of transformation-induced dislocations (Fig. 4c), and concomitant drastic particle softening (Fig. 3). All tested particles in the cycled sample exhibited continuous plastic deformation with small intermittent load drops sometimes preceded by minor strain bursts, which is typical for the small-scale samples with pre-existing dislocations [24,25,28,32]. At the first glance, this is inconsistent with the coherent metal-hydride transformation mechanism since at least some of the tested Pd particles (namely the smallest ones) should transform coherently into Pd hydride. Thus, most probably the observed dislocations formed during hydrogen desorption, in accordance with the hypothesis of Syrenova et al. [20]. At the same time, it should be noted that the nucleation of dislocations in the metal phase during hydrogen absorption cannot be excluded, as it has been shown that hydrogen dissolved in the metal phases decreases the dislocation nucleation stress in the Al-H, Ni-H [39] and V-H [40] systems. In the latter case the decrease of the nucleation stress was related to H segregation to the dislocation cores causing the decrease of their excess energy and concomitant decrease of the nucleation stress [41]. This effect should be present during hydrogen absorption and desorption and should result in dislocations in both cycle parts. However, if our experiments can be interpreted in terms of a coherent hydride precipitation during absorption, dislocation formation happens during desorption.

The fact that the size effect on strength in cycled particles is much weaker than in their as-dewetted counterparts (see Fig. 3) is closely related to the presence of dislocations in the former. Indeed, the dislocation density of $\rho = 5.7 \times 10^{14}$ $1/m^2$ measured in the cycled particles corresponds to the average distance of 42 nm between the neighboring dislocations. This distance defines an intrinsic length scale associated with cycled particles, which is significantly smaller than their external size. It has been shown in the pioneering work of Uchic et al. that the yield stress of small samples with internal dimensional scale which is much finer than their external size, is size-independent [42]. In this respect it is worth noting that extrapolating Eq. (1) for softening effect to the particle sizes larger than those studied in the present work indicates that the softening effect vanishes for the particles with $D$=2.4 μm. This is in a good agreement with the work of Chatain et al. in which a significant acceleration of equilibration kinetics of Cu particles on sapphire was observed for the particles larger than 3 μm in size [43]. This change of kinetics was attributed to a massive presence of dislocations in large as-dewetted particles. In the case of the present work this means that the dislocation density in as-dewetted particles larger than 2.4 µm in size would be comparable to that caused by hydrogen cycling.

Mordehai *et al.* reported the presence of sessile dislocations in Au nanoparticles deforming according to the nucleation-controlled plasticity mechanism [13]. These dislocations did not affect the nucleation of new dislocations at the particle-punch interface during microcompression tests, and did not cause any reduction of high compressive strength of the particles. The fact that all particles in the cycled sample exhibited continuous plastic deformation and low yield strength indicate that most dislocations introduced in the particles by hydrogen cycling are glissile.

To conclude, we demonstrated that hydrogen cycling of Pd nanoparticles fabricated by solid state dewetting method results in their drastic softening. This is opposite to the transformation strain hardening effect observed in the bulk coarse grained Pd. The hydrogen cycling also changed the mechanical response of the particles from the one typical for nucleation-controlled plasticity with elastic limit of up to 11 GPa to jerky plastic yielding typical for the samples with pre-existing dislocations and their sources. This softening effect was attributed to glissile dislocations formed in the particles due to high metal-hydride transformation strain. It is suggested that some of these dislocations which are predominately formed during hydrogen desorption, remain in the particles after hydrogen cycling. Our work demonstrates that hydrogen cycling can be employed for manipulating the mechanical properties of hydride-forming metals at the nanoscale. In particular, hydrogen cycling may help to improve the plastic deformation behavior of small pristine metal samples that is otherwise difficult to control [44] because of the stochastic nature of dislocation nucleation.

**Acknowledgements:** J.Z. and E.R. wish to thank the Israel Science Foundation (grant 617/19) for partial support of this work. E.R. wishes to thank the International Excellence Fellowship of KIT. The deposition of thin Pd films was performed at the Technion Micro- Nano- Fabrication and Printing Unit (MNFPU). C.K. and M.V. kindly acknowledge the financial support by the European Research Council (ERC) under the grant No. 101043969 (TRITIME).